\documentclass[11pt,a4paper]{article}
\usepackage[utf8]{inputenc}
\usepackage[T1]{fontenc}
\usepackage{amsmath,amssymb}
\usepackage{graphicx}
\usepackage[margin=2.4cm]{geometry}
\usepackage{hyperref}
\newcommand{\vt}{\tilde{v}}
\newcommand{\gN}{g_{\rm N}}
\newcommand{\ftrip}{f_{\rm trip}}

\title{Estimator forensics for the wide-binary gravity test: the eccentricity--triple coupling manufactures a pseudo-signal, and a pre-registered protocol for Gaia DR4}
\author{Hicham Boufourou\\
\small Independent Researcher, Brussels, Belgium --- \texttt{hicham.boufourou@hotmail.com}}
\date{\small Preprint --- August 2026}

\begin{document}
\maketitle

\begin{abstract}
Analyses of Gaia DR3 wide binaries reach opposite verdicts on a claimed low-acceleration gravitational anomaly: boost factors $\gamma=G_{\rm eff}/G\approx1.4$ (Chae 2023--2026; Hernandez et al.\ 2024) versus Newtonian gravity at high significance (Banik et al.\ 2024; Pittordis et al.\ 2025), from the same parent catalogue. We localize the disagreement experimentally. A generative population model --- Keplerian binaries plus hierarchical triples with photocentre and time-averaging effects --- is validated without tuning against the published statistics of Pittordis et al.\ (2025): triple median $\vt$ reproduced to $\leq0.04$ in all eight separation bins, cut-survival 40.8\% vs 40.5\%. Synthetic catalogues of known truth, including the intrinsic velocity truncation of the El-Badry et al.\ (2021) selection and measurement noise resampled from the real data, are then confronted with three estimator families. A full-median estimator self-calibrated on the Newtonian regime, even with a perfect eccentricity correction, recovers $\gamma=1.08$--$1.13$ from purely Newtonian universes containing 20--30\% residual triples --- half-way to the claimed anomaly --- while a tail-truncated estimator and a mixture likelihood recover $\gamma\leq1.04$ and $\gamma=1.00$. An injected MOND-like boost ($\gamma=1.4$ above 2 kau) is recovered at $\gamma\geq1.26$ by all three: the hypotheses never overlap under robust estimation, so the controversy is decidable. The exercise exposed two noise pathologies relevant to all median-based analyses: a Rice-type bias requiring zone-matched noisy templates, and parallax-noise injection when perspective corrections use individual distances. Applying the calibrated estimators to the 81,088-pair Chae (2024) sample we measure $\gamma_{\rm test}=1.045\,[1.025,1.068]$ (2--30 kau; $f(e)$ systematic band 0.96--1.05) and, from a joint $(\gamma,\ftrip)$ mixture fit, $\gamma=1.05\,[1.04,1.07]$ with $\ftrip=0.15$, consistent with the independent 0.17 of Pittordis et al.\ (2025); $\gamma=1.4$ is rejected at $\Delta\ln\mathcal L=129$ ($\approx16\sigma$). The deep-MOND bin ($\gN<0.3\,a_0$) is perspective-sensitive --- $\gamma$ moves from 0.97 to $1.11\pm0.07$ after the angular correction, partly radial-velocity noise injection --- and its full systematic bracket remains consistent with Newton while excluding 1.35--1.40. We freeze a pre-registered protocol (cuts, zones, both calibrated estimators, mandatory $f(e)$ band, decision criteria written in advance) to be time-stamped before the Gaia DR4 release, where $\sigma_{\rm stat}(\gamma)\lesssim0.005$ will make estimator systematics fully dominant.
\end{abstract}

\section{Introduction}
Wide binaries probe gravity at internal accelerations below $a_0\simeq1.2\times10^{-10}$ m s$^{-2}$, where MOND-type theories with the Galactic external field effect (EFE) predict an effective boost $\gamma\approx1.35$--$1.40$ of the gravitational constant, i.e.\ $+18\%$ on relative velocities ($\sqrt{1.4}=1.183$), while GR + dark matter predicts pure Newtonian dynamics (Banik \& Zhao 2018; Pittordis \& Sutherland 2018).

Gaia DR3 made the test statistically possible --- and produced a stark contradiction. Chae (2023, 2024), Chae et al.\ (2026) and Hernandez et al.\ (2024) report the anomaly at up to $\sim5\sigma$; Banik et al.\ (2024, $19\sigma$ for Newton), Pittordis \& Sutherland (2023), Pittordis et al.\ (2025, hereafter PS25) and the Quality Framework of Banik et al.\ (2026) find no deviation --- from the same El-Badry et al.\ (2021) parent catalogue. The disagreement is therefore purely methodological --- and the most recent exchange sharpens it: Chae \& Yoon (2026) revisit precisely the two contested levers, data quality control and multiple-star modeling, and reconfirm the anomaly from the same catalogue. Independently of the statistical route, Pasquini et al.\ (2026) find that three of twelve rigorously vetted VLT-ESPRESSO systems admit no bound Newtonian orbit, keeping the per-system route open as well. Concurrently, using a different observable --- the projected orbital momentum computed from censored astrometric parameters --- Makarov (2026) finds no manifestation of modified Newtonian dynamics in a sample of 103,169 Gaia DR3 wide binaries, providing an independent Newtonian verdict on the same data release.

Recent work has begun to localize it. On the 36-pair 3D sample (the Bayesian orbit-modeling route opened by Chae 2025), hierarchical Bayesian reanalysis of Saad \& Ting (2026) showed the answer flips between $\langle GM\rangle=1.12$ and $\langle GM\rangle=1.56$ with the deprojection choice alone. Banik et al.\ (2026) argued qualitatively that a loose upper limit on $\vt$ can generate a MOND-like signal, and PS25 criticized the calibration of the triple fraction at small separations. What has been missing is a controlled experiment: the competing estimator families confronted with the same synthetic universes of known truth, with the coupling between eccentricity modelling and tail treatment quantified. This paper provides that experiment (Sect.~4), an independent robust measurement on DR3 (Sect.~5), and a protocol frozen before Gaia DR4 (Sect.~6).

\section{Sample, observable and pipeline}
We use the public sample of Chae (2024): 81,088 pairs within 200 pc drawn from El-Badry et al.\ (2021), with proper motions and uncertainties, distances, masses, RUWE, chance-alignment probability, per-pair inferred eccentricities, and Gaia radial velocities for at least one component in 71\% of pairs. The observable is
\begin{equation}
\vt \;=\; \frac{|\Delta v_\perp|}{\sqrt{G_N M_{\rm tot}/s_{\rm 2D}}},
\end{equation}
with Newtonian normalization: an observer does not know $G_{\rm eff}$, so under $G_{\rm eff}=\gamma G_N$ the whole $\vt$ distribution is multiplied by $\sqrt{\gamma}$. Throughout, a hypothesis $\gamma$ multiplies the model $\vt$ by $\sqrt{\gamma}$ \emph{before} noise injection, the El-Badry truncation and any estimator truncation; every selection is applied to the scaled, noised distributions, identically for pseudo-data and templates, so selection effects are $\gamma$-consistent by construction.

$\Delta v_\perp$ is computed from proper-motion differences with an angular perspective correction: the systemic velocity (mean proper motion plus available systemic radial velocity) is projected at each component's position at the common mean distance, and the predicted perspective proper-motion difference is subtracted (El-Badry 2019). Two implementation traps are worth recording. First, missing radial velocities in the source catalogue are encoded as sentinels ($-10{,}000$/$-20{,}000$ km s$^{-1}$) and must be masked. Second, using individual parallax distances in the projection injects parallax noise amplified by the systemic tangential velocity ($\sim30$ km s$^{-1}$), overwhelming the $\sim0.5$ km s$^{-1}$ orbital signal; the correction must be purely angular, at the common distance. Pairs lacking any radial velocity receive the tangential part of the correction only; radial-velocity availability is 83.0\% in the validation zone, 95.7\% in the test zone and 99.4\% in the deep bin, so the tangential-only fallback operates mostly where the correction itself is negligible (validation-zone $P_{90}(|\delta\vt|)=0.0016$, versus 0.177 in the deep bin).

Pre-registered zones: validation $0.2$--$2$ kau (Newtonian regime; mandatory calibration), test $2$--$30$ kau, and a deep bin $\gN/a_0\in[0.03,0.3]$. Fiducial cuts: RUWE $<1.4$ (both components), $R_{\rm chance}<0.01$, $\sigma_{\vt}<0.10$ (41,760 pairs); robustness variants use RUWE $<1.2$ (37,197) and an HRD lobster-body proxy. A pipeline control requires the validation-zone median to match the Newtonian template to $<2\%$; we measure $+1.4\%$.

\section{Generative population model, validated on published oracles}
Binaries are Keplerian orbits with isotropic orientation, uniform-in-time phase and switchable $f(e)$ (flat; thermal; superthermal $\alpha=1.3$ of Hwang et al.\ 2022; Tokovinin \& Kiyaeva 2016; or empirical per zone). Hierarchical triples follow PS25: Kroupa present-day mass function, log-flat outer orbits, Offner et al.\ (2023) lognormal inner semi-major axes with the Tokovinin (2014) stability limit, photocentre--barycentre factor with resolved/unresolved logic at $1''$, inner velocities time-averaged over the 34-month DR3 baseline, the apparent-mass bias, and simulated RUWE/image-peak/lobster cuts.

Without any tuning, the model reproduces the published PS25 statistics, which we take as ground truth and hereafter call \emph{oracles} (i.e.\ published reference values used for validation): binary $P_{90}=0.933$--$0.941$ (oracle $0.94\pm0.01$); fractions $\vt\geq0.8$ of 23.1/21.0/20.5\% for flat/thermal/superthermal (oracle 23.2/21.0/20.6\%); triple survival of the quality cuts 40.8\% (oracle 40.5\%, RUWE rejecting 36.5\% vs $\sim37\%$); and triple median $\vt$ per separation bin from 0.87 to 1.60 versus the published 0.87--1.62, all eight bins agreeing to $\leq0.04$. Triple $P_{90}$ values run 10--14\% above PS25 --- a heavier contaminating tail, conservative for our purpose, traced to a simplified $L(M)$ relation. All synthetic catalogues include the intrinsic El-Badry velocity truncation ($\vt\leq2.23/\sqrt{M_{\rm tot}}$) and 2D measurement noise resampled pair-by-pair from the real error columns. The per-pair $\sigma_{\vt}$ is propagated treating the two components' proper-motion errors as independent; spatially correlated calibration errors are common-mode for components separated by arcminutes and largely cancel in the proper-motion \emph{difference}, so this treatment, if anything, overstates the effective noise --- an overstatement shared identically by pseudo-data and zone-matched templates.

\section{Injection--recovery: the pseudo-signal map}
We generate catalogues with truth $\gamma\in\{1.0;\ 1.4$ above 2 kau (EFE-saturated)$\}\times f(e)\in\{$thermal; superthermal; $s$-dependent$\}\times$ residual triple fraction $\ftrip\in\{0,0.1,0.2,0.3\}$, and evaluate three estimators, each given the correct per-zone $f(e)$, the same noise model and the same selection --- the only remaining degree of freedom is the treatment of the tail. E1 (full-median, Chae-like): squared ratio of full medians test/validation, corrected by the expected Newtonian ratio. E2 (truncated, this work): medians truncated at $\vt<\sqrt2$ --- the Keplerian ceiling for bound pairs, since $|\Delta v|\leq\sqrt{2GM/r}$ and $s_{\rm 2D}\leq r$ imply $\vt\leq\sqrt2$ for any bound orbit, so the excluded region can only host multiples, noise outliers or modified gravity --- with the identical truncation applied to zone-matched noisy templates, inverted on a $\gamma$ grid; the threshold choice is not delicate (Sect.~5). E3 (mixture, PS-like): Poisson likelihood of the test-zone histogram as $(1-f)\cdot{\rm binaries}_\gamma+f\cdot{\rm triple}$ template, $(\gamma,f)$ free.

\begin{figure}[!ht]
\centering
\includegraphics[width=\textwidth]{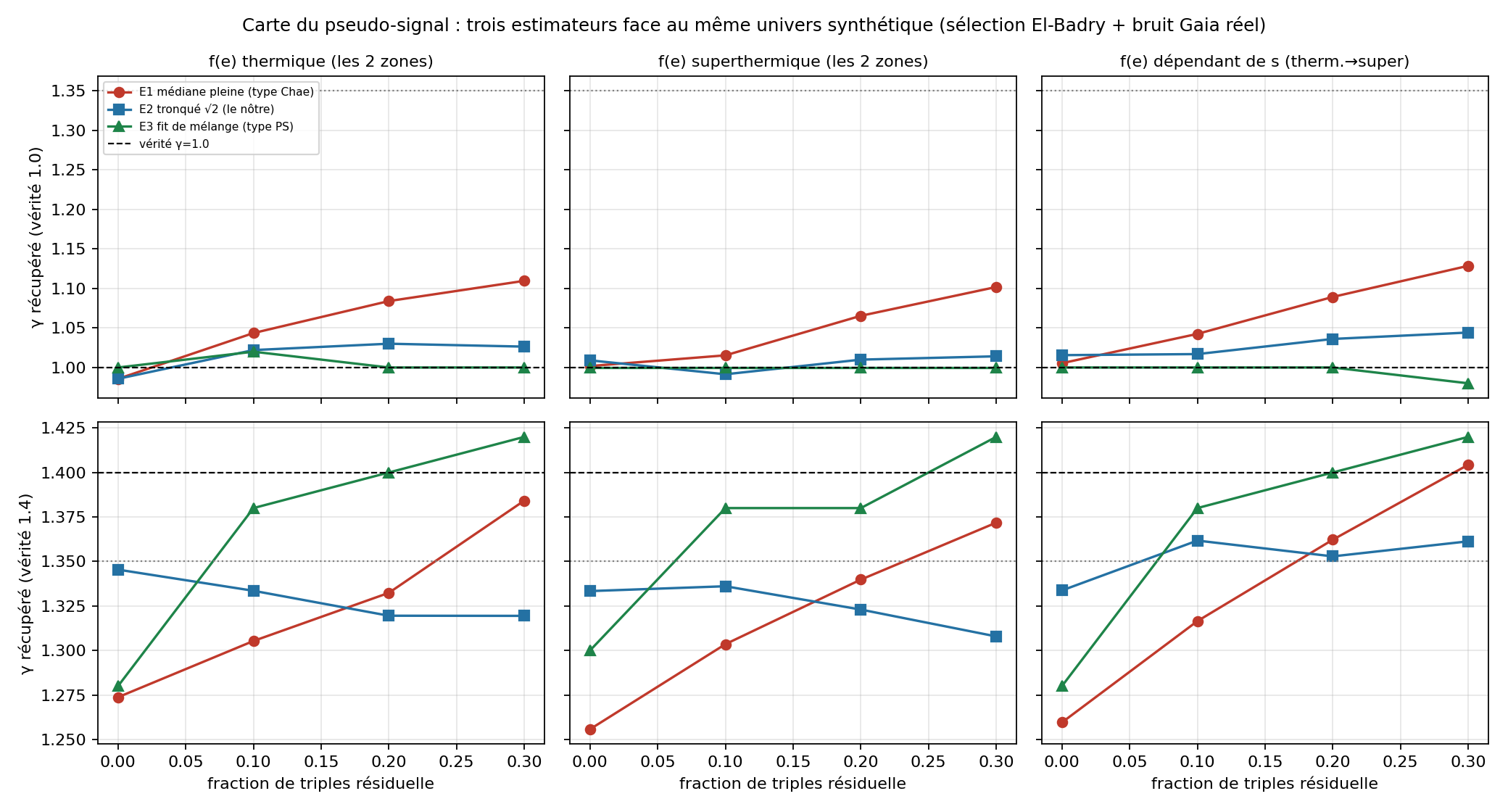}
\caption{The pseudo-signal map: three estimator families confronted with the same synthetic universes (El-Badry selection + real Gaia noise). Top row: Newtonian truth; bottom row: MOND-EFE truth ($\gamma=1.4$ above 2 kau). Columns: eccentricity distributions. On Newtonian truth the full-median estimator (red) manufactures $\gamma=1.08$--$1.13$ at realistic contamination, identically in all $f(e)$ panels; the truncated (blue) and mixture (green) estimators stay at unity. On MOND truth all three recover $\geq1.26$. Single-seed grid; statistical jitter $\approx\pm0.02$.}
\end{figure}

On Newtonian truth, E1 manufactures $\gamma=1.08$--$1.13$ at $\ftrip=0.2$--$0.3$, identically in all three $f(e)$ panels --- the culprit is the residual tail, not the eccentricity correction when the latter is correct; E2 stays $\leq1.04$ and E3 at $1.00$. On MOND truth all three recover $\geq1.26$ (E2: 1.31--1.36, the $\sim5\%$ dilution being an $a\leftrightarrow r_p$ mixing at the 2 kau boundary). The two regimes never overlap under E2/E3: the controversy is decidable with DR3-class data. Two methodological by-products: (i) a Rice-type bias --- 2D noise inflates $\|\vt\|$ more in the test zone where $v_c$ is small; before zone-matched noisy templates even E2 returned 1.18 on pure Newtonian truth; the same-cuts-on-simulations rule must extend to noise, per zone; (ii) the $\gamma\leftrightarrow\ftrip$ degeneracy of mixture fits as $\ftrip\to0$.

\section{DR3 measurement}
With the calibrated estimators: $\gamma_{\rm test}({\rm E2})=1.045\,[1.025,1.068]$ (empirical per-zone $e$), stable under RUWE $<1.2$ (1.047) and the HRD variant (1.059); $f(e)$ systematic band \{empirical 1.045; thermal 0.958; superthermal 0.957\} --- the empirical-eccentricity correction alone pushes the estimate up by $\sim9\%$, and the full band brackets unity. The result is also stable against the E2 truncation threshold: scanning $T\in[1.2,1.8]$ moves the recovered $\gamma$ by at most 0.016 (1.040 to 1.056, empirical $e$; 0.940--0.969 thermal), within the statistical interval, because data and templates are re-truncated identically at every threshold. Table~\ref{tab:dr3} collects the DR3 measurements. The mixture likelihood on the real test zone gives $\gamma=1.05\,[1.04,1.07]$ with $\ftrip=0.15$, in striking external agreement with the independent PS25 value of 0.17; $\gamma=1.4$ is rejected at $\Delta\ln\mathcal L=129$ ($\approx16\sigma$, statistical); equivalently, 1.4 lies 0.33 above the upper edge of the full $f(e)$ systematic band $[0.957,1.068]$, an order of magnitude beyond the statistical width, so the rejection is limited by neither statistics nor the eccentricity family. $\Delta\ln\mathcal L(\gamma=1)=4.8$ ($\approx3.1\sigma$, statistical) lies within that same band: the data prefer $\gamma\simeq1.05$ over 1.00 by less than the eccentricity systematic (Fig.~2).

\begin{table}[!ht]
\centering
\caption{DR3 measurements. Fiducial cuts: RUWE $<1.4$ (both components), $R_{\rm chance}<0.01$, $\sigma_{\vt}<0.10$. Intervals are 68\% bootstrap; entries without an interval are central values of robustness variants.}
\label{tab:dr3}
\small
\begin{tabular}{llllc}
\hline
Estimator & Cuts & $f(e)$ & $\gamma$ (2--30 kau) & $\gamma$ (deep bin)\\
\hline
E2 & fiducial & empirical & $1.045\,[1.025,1.068]$ & $1.11\pm0.07$\\
E2 & fiducial & thermal & 0.958 & $\approx1.01$\\
E2 & fiducial & superthermal & 0.957 & $\approx1.01$\\
E2 & RUWE $<1.2$ & empirical & 1.047 & ---\\
E2 & +HRD proxy & empirical & 1.059 & ---\\
E2, $T\in[1.2,1.8]$ & fiducial & empirical & 1.040--1.056 & ---\\
E3 (mixture) & fiducial & empirical & $1.05\,[1.04,1.07]$, $\ftrip=0.15$ & ---\\
\hline
\end{tabular}
\end{table}

\begin{figure}[!ht]
\centering
\includegraphics[width=0.95\textwidth]{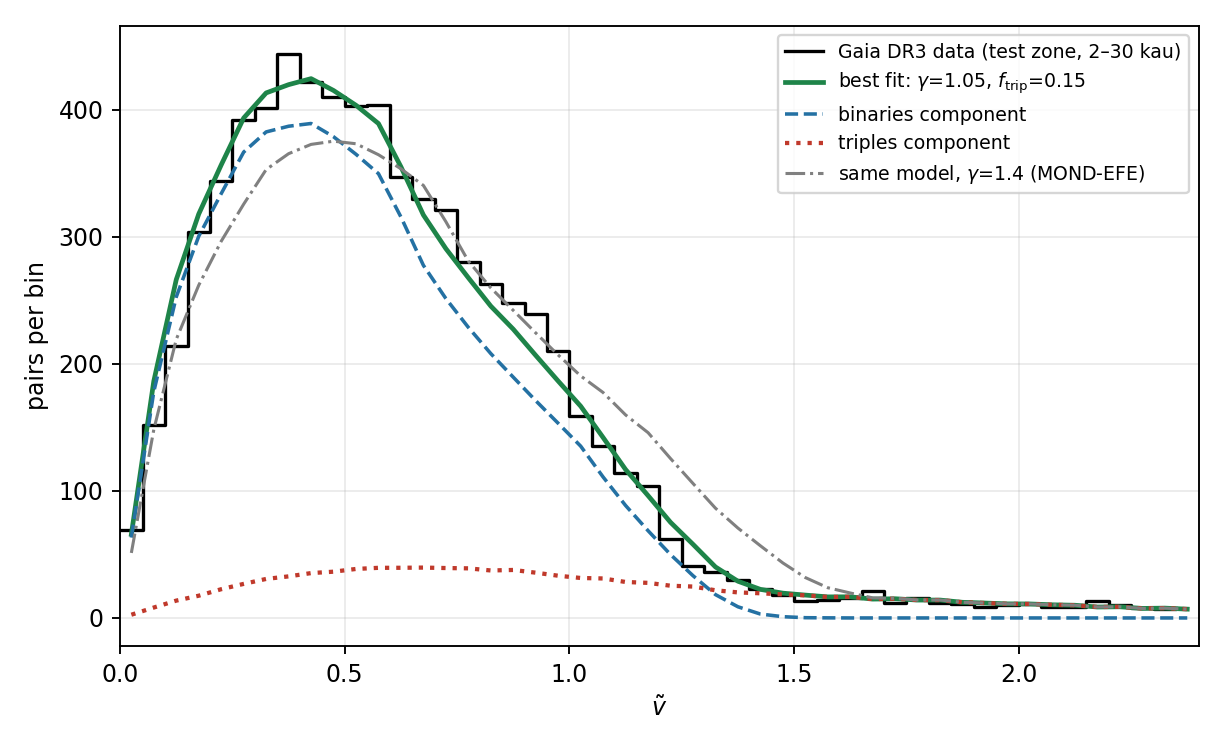}
\caption{Test-zone (2--30 kau) $\vt$ histogram of the fiducial Gaia DR3 sample versus the joint $(\gamma,\ftrip)$ mixture fit and its components. The grey dash-dotted curve keeps the best-fit $\ftrip$ and triple component fixed and rescales only the binary component to $\gamma=1.4$ (triple inner velocities are high-acceleration and carry no EFE boost); it is rejected at $\Delta\ln\mathcal L=129$.}
\end{figure}

The deep bin ($N=341$) is perspective-sensitive: $\gamma({\rm E2})=0.97$ without and $1.11\pm0.07$ with the angular correction, part of the shift being radial-velocity noise injection ($\sigma_{\rm RV}\cdot\theta$ folded into the modulus where $v_c\sim0.1$ km s$^{-1}$); with the thermal/superthermal $e$-models the corrected value drops to $\sim1.01$. The full deep-bin bracket $[\approx0.97,1.19]$ is consistent with Newton and excludes 1.35--1.40 at $\approx3\sigma$; the global test zone excludes it at $\sim16\sigma$ (statistical), with systematics bounded by the published band (Fig.~3). Reading the map of Fig.~1 backwards, a full-median analysis of this same sample with empirical-$e$ correction would report $\gamma\approx1.12$ (global) to $1.20$ (deep) --- reproducing the amplitude of the claimed anomaly from a data set that robust estimation shows to be Newtonian.

\begin{figure}[!ht]
\centering
\includegraphics[width=0.95\textwidth]{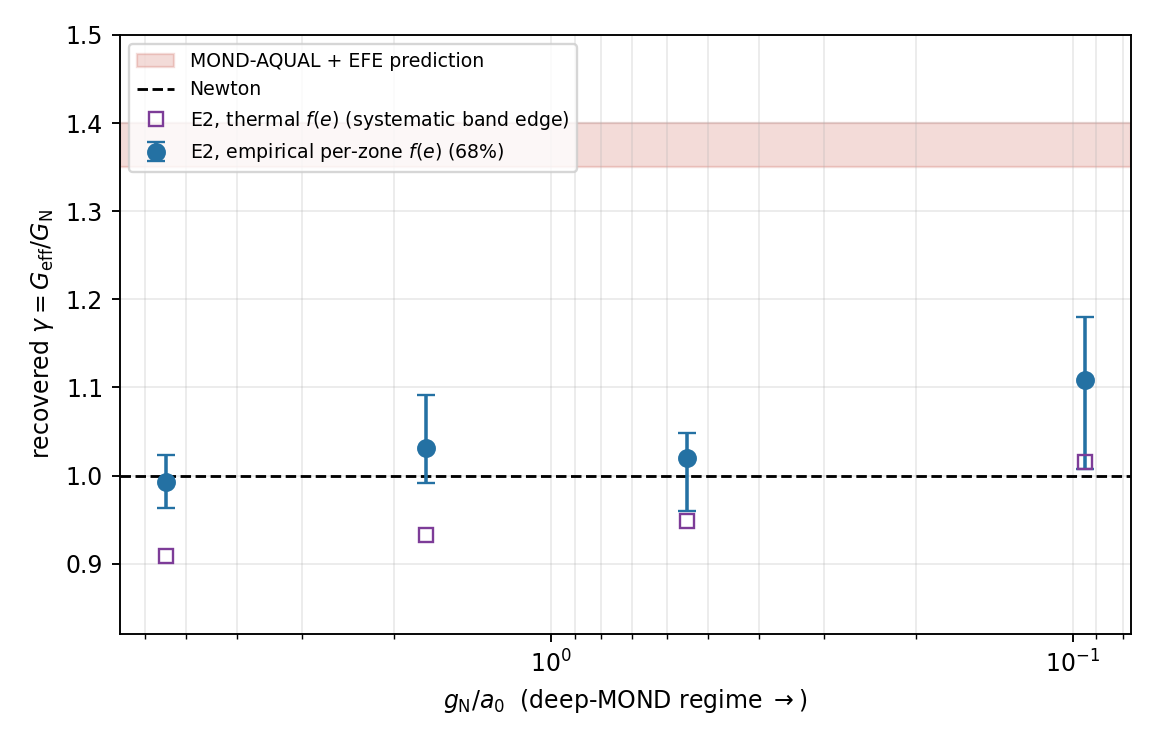}
\caption{Recovered $\gamma$ versus internal Newtonian acceleration (E2, perspective-corrected, fiducial cuts). Circles: empirical per-zone $f(e)$ with 68\% bootstrap intervals; open squares: thermal $f(e)$ (lower edge of the systematic band). The MOND-AQUAL + EFE prediction (shaded) is excluded across the full acceleration range.}
\end{figure}

\section{Pre-registered protocol for Gaia DR4}
DR4 (2026) will multiply proper-motion precision by 2--4 and the usable sample by 3--5, driving $\sigma_{\rm stat}(\gamma)$ below 0.005: estimator systematics will fully dominate, and post-hoc analysis choices will be able to produce any verdict. We therefore freeze, before the release, a complete protocol --- sample and cuts; the three zones; estimators E2 and E3 exactly as calibrated in Sect.~4 with their measured biases; the mandatory $f(e)$ band; noise-matched, selection-matched templates --- and four decision criteria written in advance, including a STOP-and-publish rule if the validation-zone control fails and an explicit grey-zone outcome. E2 is primary and E3 secondary, as fixed in the protocol; the Newton and anomaly verdicts both require the two estimators to agree, and any divergence beyond their combined intervals falls into the pre-declared grey zone, published as such. One version note: the frozen protocol (v1.0, 15 July 2026) quotes calibration-standard values from the pipeline state of that date ($\Delta\ln\mathcal L=138$, $\ftrip=0.14$, deep bin $0.97\pm0.13$); the final pipeline of this paper gives the values of Sect.~5. The protocol file is kept unchanged, as frozen; its decision criteria reference procedures and thresholds, not those standards, and are unaffected. The protocol document and code hash will be deposited on Zenodo prior to DR4; the present DR3 measurement serves as its calibration standard.

\section{Discussion}
The map quantifies both prior qualitative claims: the statement of Banik et al.\ (2026) that a loose $\vt$ limit can create a MOND-like signal, and the PS25 criticism of small-separation triple calibration, merge into a single measured number --- the leverage of the residual triple tail on a full median, $\approx+0.08$--$0.13$ in $\gamma$ at realistic contamination, insensitive to the eccentricity model when the latter is correct, and additive with $e$-model mismatch when it is not. Combined with the deprojection sensitivity demonstrated on the 3D sample by Saad \& Ting (2026), the Chae--Banik disagreement is now localized at every front where it appears. Our $\gamma=1.00$--$1.05$ also agrees with the independent Newtonian verdict that Makarov (2026) obtains from a different observable --- the projected orbital momentum --- on the same data release: two unrelated statistics, one conclusion. The deep-bin sensitivity to the perspective implementation (angular versus individual-distance projection, and the handling of missing radial velocities) is a natural candidate contributor to reported anomalies at the widest separations; we have not audited individual published pipelines, and the injection--recovery framework is the neutral ground on which any implementation can be tested. Two caveats temper the DR3 numbers rather than the conclusion: the per-pair eccentricities of Chae (2024) are prior-dependent inferences, and their use defines the upper edge of our systematic band; and the deep bin couples perspective correction, radial-velocity availability and small numbers --- a warning for any analysis that leans on the widest separations. A third caveat concerns the map itself: since the leverage of the triple tail on a full median scales, to first order, with the tail mass, the 10--14\% $P_{90}$ excess of our triple model implies the quoted $+0.08$--$0.13$ is an upper edge; matching the PS25 tail exactly would shrink it by roughly that fraction, leaving the map's topology unchanged. Our triple model, while anchored on PS25 oracles, is one family; the injection--recovery framework is precisely the tool that lets any alternative triple model be substituted and the map redrawn. If DR4 confirms $\gamma=1$ at $\sigma\approx0.005$, the wide-binary window closes for AQUAL-type external-field predictions of 1.35--1.40; whatever survives of modified dynamics would then have to decouple the binary regime from the galactic one.

\section{Conclusions}
Confronted with identical synthetic universes, the estimator families in use in the literature diverge exactly as the literature does; the divergence is manufactured by the residual-triple tail acting on full medians, not by gravity. Robust estimators calibrated on that map measure $\gamma=1.00$--$1.05$ on Gaia DR3, exclude the MOND-EFE prediction at high significance, and --- being frozen and pre-registered --- turn Gaia DR4 into a decisive, tamper-proof test.

\section*{Data and code availability}
The full pipeline (population engine, injection--recovery grid, estimators), the reduced catalogue, all validation outputs and the pre-registration document are assembled in a complete reproducibility package (scripts, data, per-phase logs) released with this preprint, archived at \url{https://github.com/HBoufourou/paperI-wide-binaries} and Zenodo DOI \url{https://doi.org/10.5281/zenodo.22073431}.

\section*{Declaration on computational tools}
This work was carried out by the author alone, without institutional affiliation. Large language models were used as computational and analytical assistants throughout: to write and debug the numerical scripts, to run the scans and minimisations, to produce the figures, to check algebra and dimensional consistency, to search and summarise the literature, and to audit drafts for internal contradictions. Several distinct systems were used and cross-checked against one another; no single system's output was accepted without an independent numerical or analytical verification.

The author is solely responsible for the physical hypotheses, the interpretation of every result, the epistemic labels attached to each claim, and the decision of what to publish and what to withdraw. All quantitative statements in this paper are reproduced by the scripts of the reproducibility package; a reader who runs them obtains the numbers printed here without needing any language model. Where an earlier draft contained an error introduced by an unverified computation, it is recorded as a withdrawal rather than silently removed.

\end{document}